\documentclass[graybox]{svmult}
\usepackage{amsfonts}
\usepackage{amssymb}
\usepackage{amsmath}
\usepackage{eucal}
\usepackage{mathrsfs}
\allowdisplaybreaks[0]
\usepackage{bm}
\usepackage{graphicx}
\usepackage{caption}
\usepackage{subcaption}
\usepackage{algorithm,algorithmic}
\usepackage{framed}
\usepackage{diagbox}
\usepackage[titletoc,title]{appendix}
\usepackage{tikz}
\usepackage{color}

\usepackage{hyperref}
\usepackage{cleveref}
\usepackage{autonum}

\usepackage{type1cm}        
\usepackage{makeidx}         
\usepackage{graphicx}        
\usepackage{multicol}        
\usepackage[bottom]{footmisc}

\usepackage{newtxtext}       %
\usepackage{newtxmath}       

\makeindex             

\begin{document}

\title*{Space-adaptive anisotropic bivariate Laplacian regularization for image restoration}
\author{Luca Calatroni, Alessandro Lanza, Monica Pragliola and Fiorella Sgallari}
\institute{Luca Calatroni \at CMAP, CNRS, \'Ecole Polytechnique, Institut Polytechnique de Paris, Palaiseau, 91128, Route de Saclay, France  \email{luca.calatroni@polytechnique.edu}
\and Alessandro Lanza \at University of Bologna, Piazza di Porta San Donato 5, Bologna, Italy \email{alessandro.lanza2@unibo.it} \and Monica Pragliola \at University of Bologna, Piazza di Porta San Donato 5, Bologna, Italy \email{monica.pragliola2@unibo.it} \and Fiorella Sgallari \at University of Bologna, Piazza di Porta San Donato 5, Bologna, Italy \email{fiorella.sgallari@unibo.it}}
%
%
\maketitle

\abstract*{In this paper we present a new regularization term for variational image restoration which can be regarded as a  space-variant anisotropic extension of the classical isotropic Total Variation (TV) regularizer. The proposed regularizer comes from the statistical assumption that the gradients of the target image distribute locally according to space-variant bivariate Laplacian distributions. The highly flexible variational structure of the corresponding regularizer encodes several free parameters which hold the potential for faithfully modelling the local geometry in the image and describing local orientation preferences.
For an automatic estimation of such parameters, we design a robust maximum likelihood approach and report results on its reliability on synthetic data and natural images. 
A minimization algorithm based on the Alternating Direction Method of Multipliers (ADMM) is presented for the efficient numerical solution of the proposed variational model. Some experimental results are reported which demonstrate the high-quality of restorations achievable by the proposed model, in particular with respect to classical Total Variation regularization.}

\abstract{In this paper we present a new regularization term for variational image restoration which can be regarded as a  space-variant anistropic extension of the classical Total Variation (TV) regularizer. The proposed regularizer comes from the statistical assumption that the gradients of the unknown target image distribute locally according to space-variant bivariate Laplacian distributions. The high flexibility of the proposed regularizer holds the potential for effectively modelling local image properties, in particular driving in an adaptive manner the strength and direction of non-linear TV-diffusion.
The free parameters of the regularizer are automatically set - and, eventually, updated - based on a robust Maximumum Likelihood estimation procedure. 
A minimization algorithm based on the Alternating Direction Method of Multipliers is presented for the efficient numerical solution of the proposed variational model. Some experimental results are reported which demonstrate the high-quality of restorations achievable by the proposed model, in particular with respect to classical TV-regularized models.}
\vspace{0.5cm}
\section{Introduction}
\label{sec:intro}
\begin{figure}[tbh]
	\center
	\begin{tabular}{cccc}
		\includegraphics[width=1.1in]{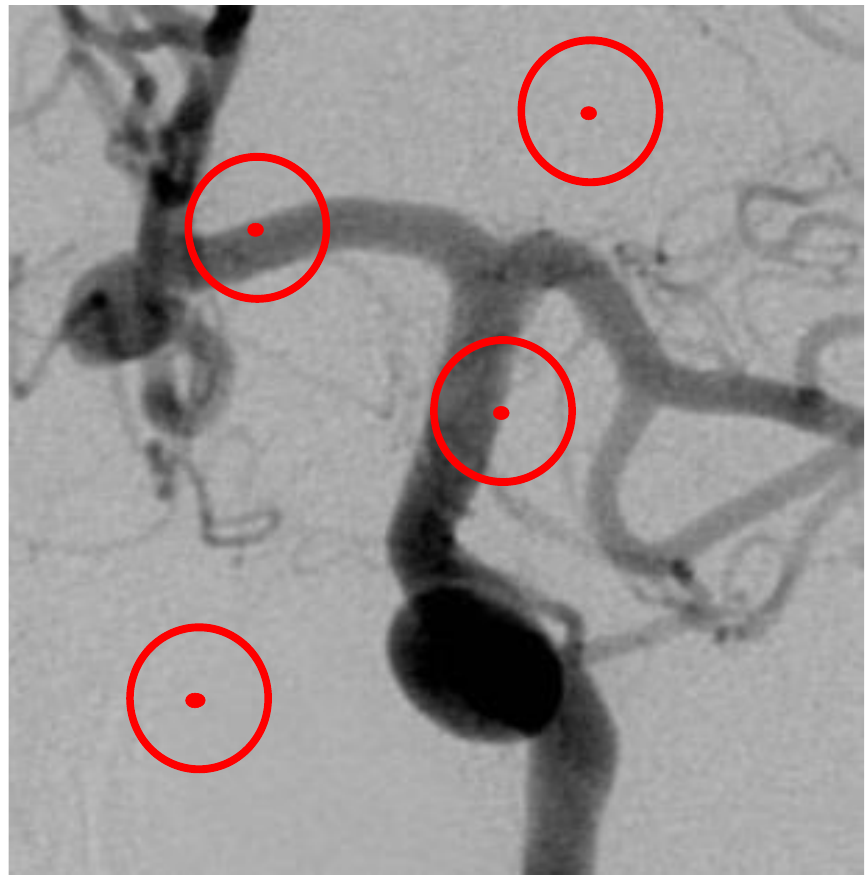} &
		\includegraphics[width=1.1in]{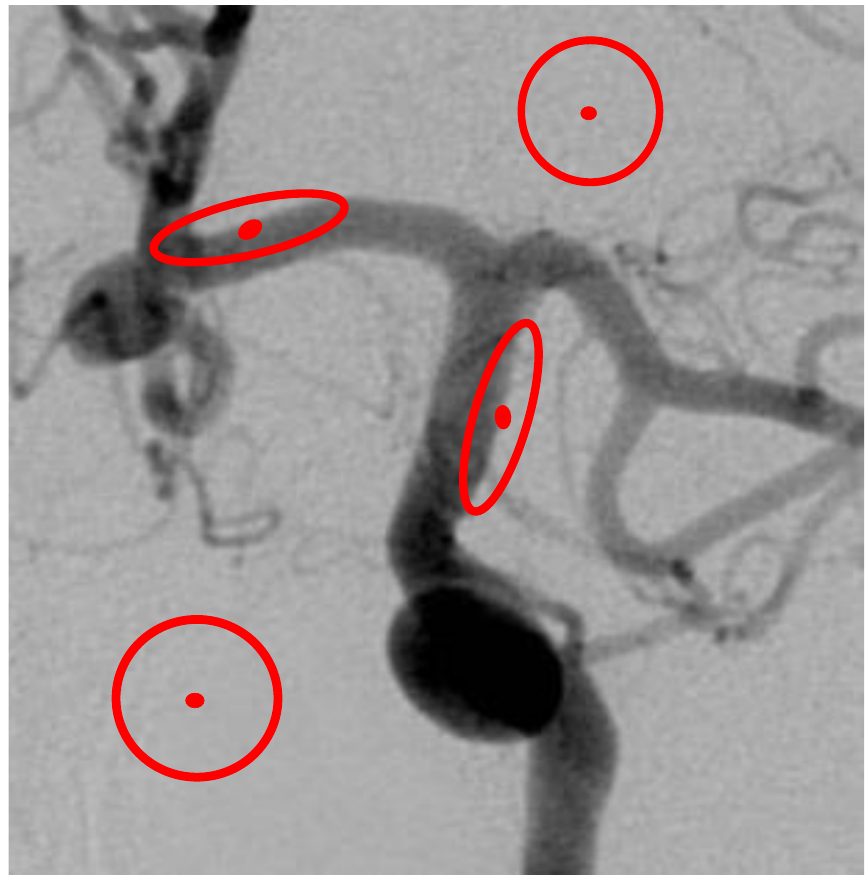}&
			\includegraphics[width=1.1in]{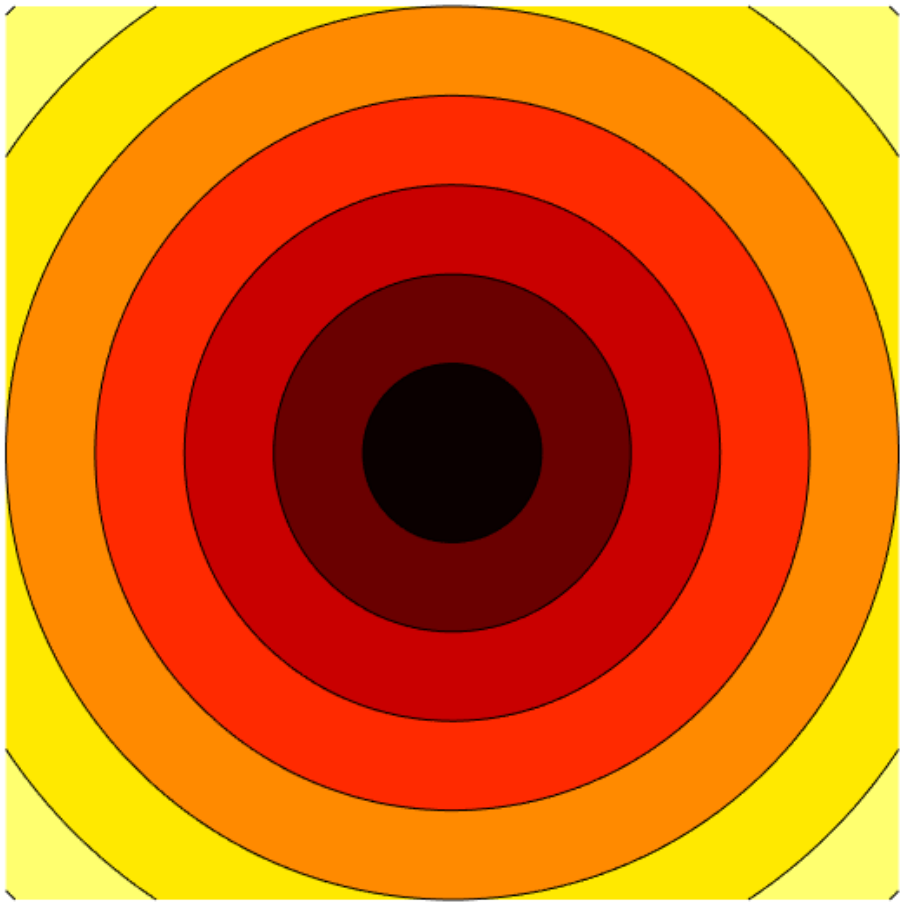} &
		\includegraphics[width=1.1in]{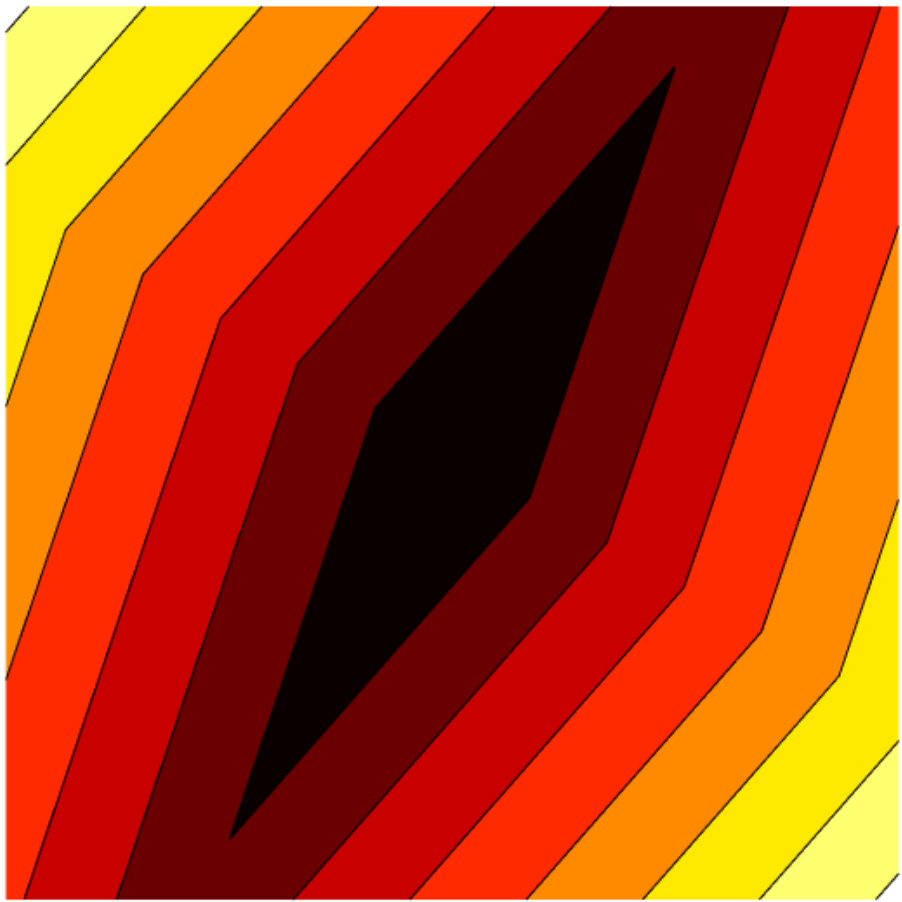}\\
		(a)&(b)&(c)&(d)\\
	\end{tabular}		
	\caption{Space-invariant isotropic TV-diffusion (a), space-variant anisotropic BLTV-diffusion (b), level curves of TV (c) and BLTV (d) regularization terms.}
	\label{fig:pdf}
\end{figure}

Image restoration is the task of recovering a sharp image $u$ starting from a blurred and noisy observation $g$. In this work, we consider a degradation model of the form
\begin{equation}
\label{eq:degmod}
g = Ku + e\,, 
\end{equation}
where $g,u\in\mathbb{R}^n$ are vectorized images, $K \in \mathbb{R}^{n\times n}$ is the linear blur operator and $e\in\mathbb{R}^n$ is an additive noise vector. A possible strategy to overcome the ill-posedeness of the linear system in \eqref{eq:degmod} is to reformulate the problem. We thus look for $u^*$, estimate of the original $u$, which solves a well-posed problem. In the \emph{variational} approach, $u^*$ is the minimizer of a cost functional $\mathcal{J}(u;K,g):\mathbb{R}^n\to \mathbb{R}$. In formula,
\begin{equation}
\label{eq:varmod}
u^*\in\arg\min_{u\in\mathbb{R}^n} \; \big\{\mathcal{J}(u;K,g):=\mathcal{R}(u) + \mu \mathcal{F}(u;K,g)\big\}.
\end{equation}
The functionals $\mathcal{R}$ and $\mathcal{F}$ are commonly referred to as the \emph{regularization} and the \emph{data fidelity} term, respectively. While $\mathcal{R}$ encodes prior information on the desired image $u$, $\mathcal{F}$ is a data term which measures the `distance' between the given image $g$ and $u$ after the action of the operator $K$ with respect to some norm corresponding to the noise statistics in the data, cf., e.g., \cite{stuart}. The regularization parameter $\mu > 0$ controls the trade-off between the two terms.
\\In this paper, we consider an Additive White Gaussian Noise (AWGN) corrupting the blurred image $Ku$, i.e. $e\sim\mathcal{N}(0,\sigma^2I_n)$, where $I_n$ is the $n$-dimensional identity matrix.
It is well known that, in presence of AWGN, a suitable choice for $\mathcal{F}(u;K,g)$ is the so-called L$_2$ fidelity term, reading as,
\begin{equation}
\label{eq:l2}
\mathcal{F}(u;K,g) = \mathrm{L}_2(u;K,g) = \frac{1}{2} \lVert Ku - g \rVert_2^2.
\end{equation}
A popular choice for the regularization term $\mathcal{R}(u)$ is given by the TV semi-norm \cite{ROF},
\begin{equation}
\mathcal{R}(u) = \mathrm{TV}(u)=\sum_{i=1}^{n}  \lVert (\nabla u)_{i} \rVert_2 \, ,
\label{eq:TV}
\end{equation}
where $(\nabla u)_i := \big( (D_h u)_i , (D_v u)_i \big)^T \in \mathbb{R}^2$ denotes the discrete gradient of image $u$ at pixel $i$,
with $D_h,D_v \in \mathbb{R}^{n \times n}$ linear operators representing finite difference discretizations of the first-order horizontal and vertical partial derivatives, respectively. 
\\Coupling the L$_2$ data term with the TV regularizer leads to one of the most widespread variational models for image restoration problem, the TV-L$_2$ (or ROF) model,
\begin{equation}
\label{eq:tvrof}
u^*\in\arg\min_{u\in\mathbb{R}^n} \bigg\{\sum_{i=1}^{n}  \lVert (\nabla u)_{i} \rVert_2 +
\frac{\mu}{2} \lVert Ku - g \rVert_2^2\bigg\}.
\end{equation}
The global perspective of the TV-L$_2$ model does not allow to diversify the action of the regularizer on regions of the image presenting different properties. In \cite{siam,ncmip,cmbbe,vip}, the authors have proposed space-variant regularization term based on statistical assumptions on the distribution of the $\ell_2$-norm of the gradients and on the gradients themselves.
\\In this paper, we propose a space-variant anisotropic extension of the TV regularizer in (\ref{eq:TV}) which, as it will be illustrated in Sect. \ref{sec:map}, comes from the \emph{a priori} assumption that the gradients of the target image $u$ distribute locally according to a Bivariate Laplace Distribution (BLD). The proposed BLTV regularizer takes the form
\begin{eqnarray}
\label{eq:jlapreg}
\mathrm{BLTV}(u;\lambda_1,\lambda_2,\theta) &=& \sum_{i=1}^{n} \lVert \Lambda_i R_{\theta_i} (\nabla u)_i\rVert_1\\
&=& \sum_{i=1}^{n}\bigg[\lambda_i^{(1)}|\langle r_i^{(1)},(\nabla u)_i \rangle| + \lambda_i^{(2)} |\langle r_i^{(2)},(\nabla u)_i \rangle|\bigg]\,,
\end{eqnarray}
where $\Lambda_i$ is a $2\times 2$ positive definite diagonal matrix and $R_{\theta_i}$ is the rotation matrix corresponding to the angle $-\theta_i$. Mathematically,
\begin{equation}
\label{eq:lamrth}
\Lambda_i = \begin{pmatrix}
\lambda_i^{(1)} & 0 \\
0 &\lambda_i^{(2)}
\end{pmatrix}\,,\quad
R_{\theta_i} = \begin{pmatrix}
\cos \theta_i & \sin \theta_i \\
-\sin \theta_i &\cos \theta_i
\end{pmatrix} = 
\begin{pmatrix}
r_i^{(1)}\\
r_i^{(2)}
\end{pmatrix}.
\end{equation}
We denote by $\lambda_1,\lambda_2,\theta \in \mathbb{R}^n$ the maps of the parameters defining the local distributions. Hence, the proposed BLTV-L$_2$ variational restoration model reads as
\begin{equation}
\label{eq:jtvl2}
u^*\in\arg\min_{u\in\mathbb{R}^n} \bigg\{\mathrm{BLTV}(u;\lambda_1,\lambda_2,\theta)+
\frac{\mu}{2} \lVert Ku - g \rVert_2^2\bigg\}.
\end{equation}
The $3n$ free parameters defining the BLTV regularizer in (\ref{eq:jlapreg})--(\ref{eq:lamrth}) hold the potential for effectively modelling local image properties, in particular driving in a suitable adaptive manner the strength and direction of non-linear TV-diffusion.
In Figs.\ref{fig:pdf}(a)-(b) the red ellipses represent  TV-diffusion strengths along all possible directions at few sample pixel locations for TV and BLTV regularizers. It is evident how for classical TV such ellipses turn out to be circles (isotropy) of constant radius (space-invariance), whereas our BLTV regularizer allows for  ellipses (anistropy) of different size (space-variance).
In practice, such flexibility of BLTV can be (and will be) exploited to diffuse in different ways in regions exhibiting different properties: for instance, strong isotropic diffusion in homogeneous regions, strongly anisotropic diffusion in regions characterized by a dominant edge direction.
In Figs.\ref{fig:pdf} (c)-(d) we show the level curves of the TV regularizer and one among the infinity of possible configuration of the level curves of the BLTV regularizer, respectively, revealing once again the flexibility of the proposed regularizer.

Together with the variational model in \eqref{eq:jtvl2}, we also propose an efficient and robust procedure for automatically estimating the parameter maps $\lambda^{(1)}$, $\lambda^{(2)}$, $\theta$ based on a Maximum Likelihood (ML) approach. The parameter maps can be updated along the iterations of the Alternating Direction Methods of Multipliers (ADMM), which is the algorithm adopted here to solve the minimization problem. Notice that the convexity of the BLTV-L$_2$ model ensures the convergence of the ADMM.

\vspace{-0.5cm}

\section{Deriving the model via MAP}
\label{sec:map}
Applying the \emph{Maximum A Posteriori} (MAP) estimation approach to image restoration consists in computing the restored image as a global maximizer of the posterior probability Pr$(u|g;K)$ of the unknown target image $u$ given the observation $g$, namely: 
\begin{equation}
u^* \:\;{\in}\;\: \arg \max_{u \in \mathbb{R}^n}\;
\text{Pr}(u|g;K) = \arg \min_{u \in \mathbb{R}^n} \;
\left\{ \,-\mathrm{ln} \text{Pr}(g|u;K)-\mathrm{ln} \text{Pr}(u)
\, \right\}\,,
\label{eq:map1}
\end{equation}
where, after applying the Bayes' rule, we drop the \emph{evidence} term Pr$(g)$ and extracted $-\ln$ of the objective function. The two terms Pr$(u)$ and Pr$(g|u;K)$ in 
(\ref{eq:map1}) are referred to as the \emph{prior} and the \emph{likelihood}.
The likelihood term associated with AWGN corruption takes the form 
\begin{equation}
\mathrm{Pr}(g|u;K)=\prod_{i=1}^{n}\,\frac{1}{\sqrt{2\pi}\sigma}\,\text{exp}\bigg(-\frac{(Ku-g)_{i}^{2}}{2\sigma^{2}}\,\bigg)=W\,\text{exp}\bigg(-\frac{\lVert Ku-g\rVert_{2}^{2}}{2\sigma^{2}}\,\bigg),
\label{eq:g_like}
\end{equation}
where $\sigma>0$ denotes the AWGN standard deviation and $W > 0$ is a normalization constant.
For what concerns the prior, a common choice is to model the unknown image $u$ as a Markov Random Field (MRF) such that the image can be characterized by its Gibbs prior distribution, whose general form is:
\begin{equation}
\text{Pr}(u)= \prod_{i = 1}^{n} \text{Pr}_i(u) = \prod_{i = 1}^{n} z_i \, \text{exp}\,(\,-\alpha \, V_{c_{i}}(u)\,)=  Z \, \text{exp}\,\bigg(\,-\alpha\,\sum_{i = 1}^{n}  V_{c_{i}}(u)\,\bigg),
\label{eq:mrf}
\end{equation}
where $\alpha>0$ is the MRF parameter, $\{c_{i}\}_{i=1}^{n}$ is the set of all cliques (a clique is a set
of neighboring pixels) for the MRF, $V_{c_{i}}$ is the potential function defined on the clique $c_{i}$
and $Z = \prod_{i = 1}^{n} z_i$ is a normalization constant. Choosing as potential function at the generic $i$-th pixel the magnitude of the discrete gradient at the same pixel, i.e. $V_{c_{i}}=\lVert (\nabla u)_{i} \rVert_{2}$ for any $i \in \{1,\ldots,n\}$, the Gibbs prior in \eqref{eq:mrf} reduces to the TV prior which, plugged into (\ref{eq:map1}), yields the popular TV regularizer. We remark that the TV prior corresponds to choosing 
\begin{equation}
\label{eq:Pri_TV}
\text{Pr}_i(u) = z_i \, \text{exp}\,\left(\,-\alpha \lVert (\nabla u)_{i} \rVert_{2} \right)\, . 
\end{equation}
Here, we propose a space-variant anisotropic generalization of the local prior $\text{Pr}_i(u)$ above. More in detail, we assume that the discrete gradient at any pixel $i$ distributes according to a space-variant BLD, such that our prior reads as 
\begin{eqnarray}
\notag
\mathrm{Pr}(u) &=& \prod_{i = 1}^{n} \text{Pr}_i(u) = 
\prod_{i = 1}^{n} z_i \, 
\exp\bigg(- \lVert \Lambda_iR_{\theta_i} (\nabla u)_i \rVert_1\bigg)\\ &=& Z \, \exp\bigg(-\sum_{i=1}^{n}\lVert \Lambda_iR_{\theta_i} (\nabla u)_i \rVert_1\bigg)\,,
\label{eq:jtvpr}
\end{eqnarray} 
where $Z$ is the normalization constant and $\Lambda_i, R_{{\theta}_i} \in \mathbb{R}^{2 \times 2}$ are defined in (\ref{eq:lamrth}).

Plugging the prior \eqref{eq:jtvpr} and the likelihood \eqref{eq:g_like} into \eqref{eq:map1} and neglecting the  constant terms, the proposed BLTV-L$_2$ variational model \eqref{eq:jtvl2} is obtained.

\vspace{-0.5cm}

\section{Parameter estimation via ML}
\label{sec:ml}
In order to make the introduction of the proposed regularizer actually useful, an efficient, robust, and automatic procedure for the estimation of the  parameter maps $\lambda^{(1)},\lambda^{(2)},\theta\in \mathbb{R}^n$ identifying all the local BLDs has to be proposed as well. To this aim, we resort to the ML approach. Consider a set of $N$ 2-dimensional samples 
$\mathcal{S} := \left\{ s_1,\ldots,s_N \right\}$ drawn from a BLD with parameters $(\lambda_1,\lambda_2,\theta)$. Here, the samples play the role of image gradients at pixels of a neighborhood of radius $r$ centered at a generic pixel $i$.
Assuming independence of the samples, according to the definition of BLD, the likelihood function is defined by
\begin{equation}
 \label{eq:ml}
 \mathrm{Pr}(\mathcal{S}|\lambda_1,\lambda_2,\theta) = \bigg(\frac{\lambda_1\,\lambda_2}{4}\bigg)^N \exp\bigg(-\sum_{i=1}^{N}\bigg(\lambda_1|\langle r_1, s_i\rangle|+\lambda_2|\langle r_2, s_i\rangle|\bigg)\bigg)\,,
 \end{equation}
where, clearly, $r_1$ and $r_2$ depend on $\theta$ - see \eqref{eq:lamrth}.
The goal here is to find 
\begin{eqnarray}
\notag
(\lambda_1^*,\lambda_2^*,\theta^*) &\in& \arg\max_{\lambda_1,\lambda_2,\theta} \mathrm{Pr}(\mathcal{S}|\lambda_1,\lambda_2,\theta) = \arg\min_{\lambda_1,\lambda_2,\theta}  -\mathrm{ln} \mathrm{Pr}(\mathcal{S}|\lambda_1,\lambda_2,\theta)\\
&=&-N\mathrm{ln}\frac{\lambda_1\,\lambda_2}{4} + \sum_{i = 1}^{N}\bigg(\lambda_1|\langle r_1, s_i\rangle|+\lambda_2|\langle r_2, s_i\rangle|\bigg). 
\label{eq:obj_ml}
\end{eqnarray}
Imposing a first-order optimality condition with respect to $\lambda_1,\lambda_2$ leads to the following closed-form estimation formulas:
\begin{equation}
\label{eq:updm}
\lambda_1 = \bigg(\frac{1}{N}\sum_{i = 1}^{N}|\langle r_1, s_i\rangle|\bigg)^{-1}\,,\quad
\lambda_2 = \bigg(\frac{1}{N}\sum_{i = 1}^{N}|\langle r_2, s_i\rangle|\bigg)^{-1}.
\end{equation}
Substituting the expressions in \eqref{eq:updm} into the objective function \eqref{eq:obj_ml}, we thus obtain the simplified minimization problem in the only variable $\theta$:
\begin{equation}  \label{pb:min_theta}
\theta^*\in \arg \min_{\theta}\bigg\{\mathrm{ln}\bigg(\sum_{i=1}^{N}|\langle r_1, s_i\rangle|\bigg)+\mathrm{ln}\bigg(\sum_{i=1}^{N}|\langle r_2, s_i\rangle|\bigg)\bigg\}\,.
\end{equation}

\vspace{-0.5cm}

\section{ADMM}
\label{sec:admm}
In order to solve numerically the proposed image restoration model \eqref{eq:jtvl2}, we use an ADMM-based algorithm - see \cite{boyd}. We first introduce two auxiliary variables $w \in \mathbb{R}^n$, $t \in \mathbb{R}^{2n}$ and rewrite the model in the equivalent linearly constrained form:

\vspace{-0.65cm}
\begin{eqnarray}\label{eq:PM_ADMM_a}
\{ \, u^*,w^*,t^* \}&
\:\;{\in}\;\:&
\arg\min_{u,w,t}
\bigg\{ \:
\sum_{i = 1}^{n}\lVert \Lambda_i R_{\theta_i} t_i\rVert_1
\;{+}\;
\frac{\mu}{2} \, \| w \|_2^2
\: \bigg\}\\ 
&&\mathrm{subject}\;\mathrm{to}:\quad
w \;{=}\; K u - g, \;\:
t \;{=}\; Du ,
\end{eqnarray}
where the space-variant matrices $\Lambda_i$, $R_{\theta_i}$ can be estimated via the ML procedure described in Sect. \ref{sec:ml} based only on the observed image $g$ (i.e. as a preliminary pre-processing step) or also updated along the ADMM iterations. 
We define the augmented Lagrangian functional:
\begin{eqnarray}\nonumber
\mathcal{L}(u,w,t;\rho_w,\rho_t)&:=& \sum_{i = 1}^{n}\lVert \Lambda_i R_{\theta_i} t_i \rVert_1+\frac{\mu}{2} \| w \|_2^2 - \rho_t^{T}(t - D u) +\frac{\beta_t}{2} \| t - D u \|_2^2\\
\label{eq:PM_AL}
&-&\rho_w^{T}(w - (Ku-g))+ \frac{\beta_w}{2}\| w - (Ku-g) \|_2^2,
\end{eqnarray}
where $\beta_w, \beta_t > 0$ are scalar penalty parameters and $\rho_w \in \mathbb{R}^n$, $\rho_t \in \mathbb{R}^{2n}$
are the vectors of Lagrange multipliers associated with the given linear constraints. 
The solution $\{u^*,w^*,t^*\}$ of problem (\ref{eq:PM_ADMM_a}) is a saddle point for $\mathcal{L}$ in (\ref{eq:PM_AL}), 
{see, e.g., \cite{boyd}}. Hence, we can alternate a minimization step with respect to $t,u,w$ with a maximization step with respect to $\rho_t,\rho_w$. Mathematically,

\begin{eqnarray}
&
u^{(k+1)} &
\;{\leftarrow}\;\;\,\,
\arg\min_{u \in \mathbb{R}^n} \;
\mathcal{L}(u,w^{(k)},t^{(k)};\rho_w^{(k)},\rho_t^{(k)}) \, ,
\label{eq:PM_ADMM_u} \\
&
w^{(k+1)} &
\;{\leftarrow}\;\;\,\,
\arg\min_{r \in \mathbb{R}^n} \;
\mathcal{L}(u^{(k+1)},w,t^{(k)};\rho_w^{(k)},\rho_t^{(k)}) \, ,
\label{eq:PM_ADMM_r} \\
&
t^{(k+1)} &
\;{\leftarrow}\;\;\,\,
\arg\min_{t \in \mathbb{R}^{2n}} \;
\mathcal{L}(u^{(k+1)},w^{(k+1)},t;\rho_w^{(k)},\rho_t^{(k)}) \, ,
\label{eq:PM_ADMM_t} \\
&
\rho_w^{(k+1)} &
\;{\leftarrow}\;\;\,\,
\rho_w^{(k)} \;{-}\; \beta_r \, \big( \, w^{(k+1)} \;{-}\; (K u^{(k+1)}-g) \, \big) \, ,
\label{eq:PM_ADMM_lr} \\
&
\rho_t^{(k+1)} &
\;{\leftarrow}\;\;\,\,
\rho_t^{(k)} \;{-}\; \beta_t \, \big( \, t^{(k+1)} \;{-}\; D u^{(k+1)} \, \big) \, .
\label{eq:PM_ADMM_lt}
\end{eqnarray}

\noindent {The solution of the primal sub-problem \eqref{eq:PM_ADMM_u} can be efficiently computed by means of standard linear Fast Fourier Transform (FFT) solvers. The sub-problem \eqref{eq:PM_ADMM_r} can be solved in closed-form by following \cite[Section 3]{ncmip}. Finally, the sub-problem \eqref{eq:PM_ADMM_t} can be solved by computing efficiently the proximal operator of the anisotropic $1$ norm, for which the proof of \cite[Proposition 6.3]{siam} can be easily adapted. The regularization parameter $\mu$ is updated along the iterations so as to fulfill the
global discrepancy principle as described in \cite{ape}. We refer the reader also to \cite{cmbbe,tvp} for more details on the numerical solution of the algorithm.}

\vspace{-0.5cm}

\section{Experimental results}
\label{sec:test}
In this section, we evaluate the performance of the proposed BLTV-L$_2$ restoration model compared with the baseline TV-L$_2$ model, also solved by ADMM. The stopping criteria of the ADMM for both models are defined based on the number of iterations as well as on the iterates relative change, i.e. we stop iterating as soon as
\begin{equation}
k \geq 1500 \;\;\; \mathrm{or}\;\;\; \delta^{(k)}:= \frac{\lVert u^{(k)} - u^{(k-1)} \rVert}{\lVert u^{(k-1)}\rVert} \geq 10^{-6}.
\end{equation}
The quality of the restored images $u^*$ is measured by means of the Improved Signal-to-Noise Ratio $\mathrm{ISNR}(u^*,g,u) = 10\log_{10}\frac{\|g-u\|_2^2}{\|u^*-u\|_2^2}$, with $u$ denoting the original uncorrupted image, and of the Structural-Similarity-Index (SSIM) \cite{ssim}.\\

\begin{figure}[tbh]
	\center
	\begin{tabular}{cccc}
		\includegraphics[width=1.1in]{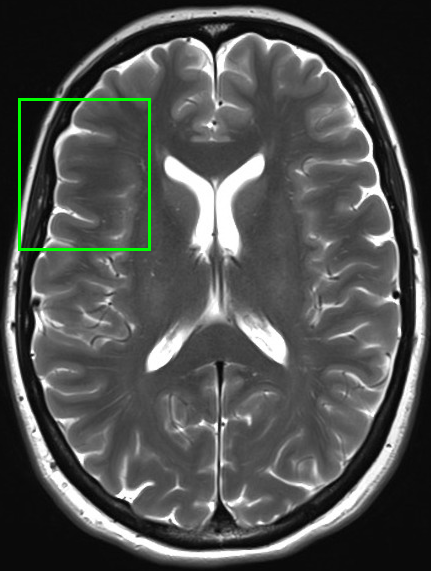} &
		\includegraphics[width=1.1in]{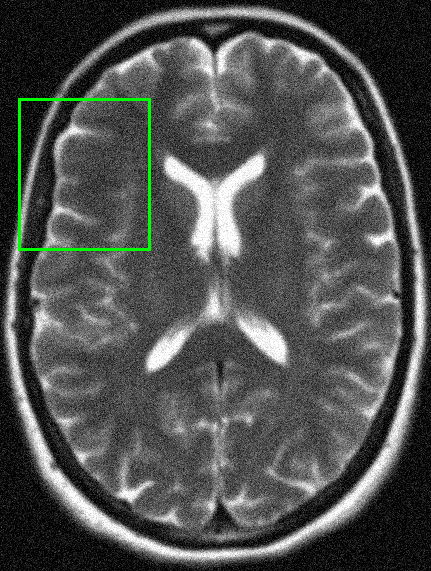} &
		\includegraphics[width=1.1in]{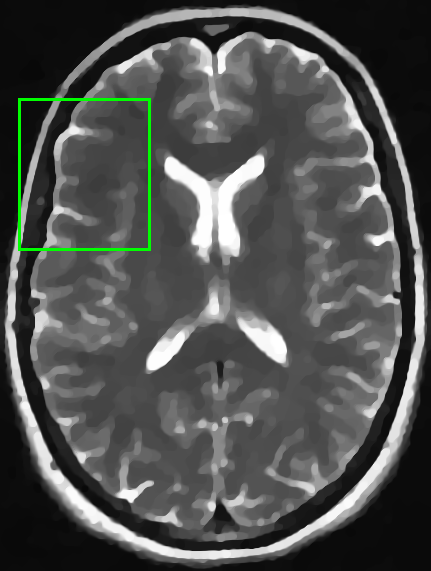} &
		\includegraphics[width=1.1in]{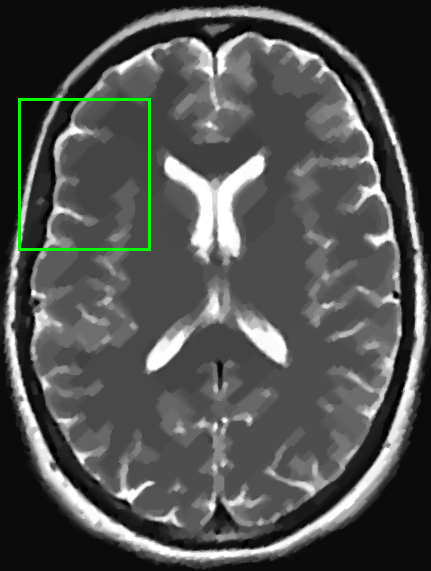}\\
		(a)&(b)&(c)&(d)\\
		\includegraphics[width=1.1in]{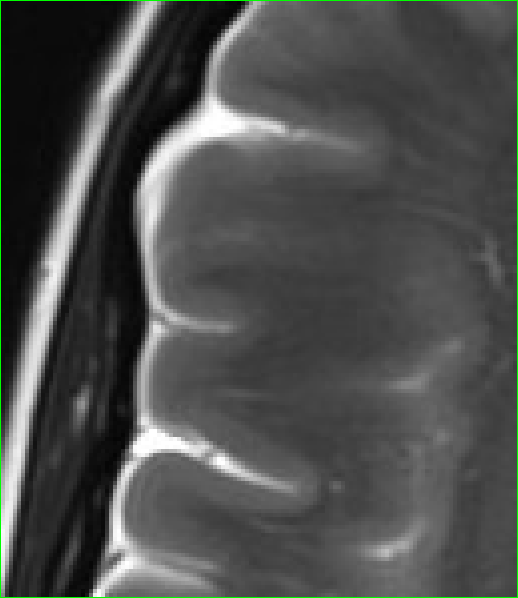} &
		\includegraphics[width=1.1in]{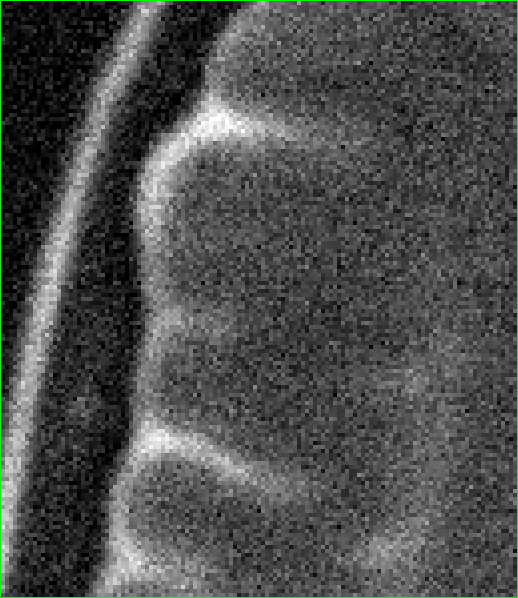} &
		\includegraphics[width=1.1in]{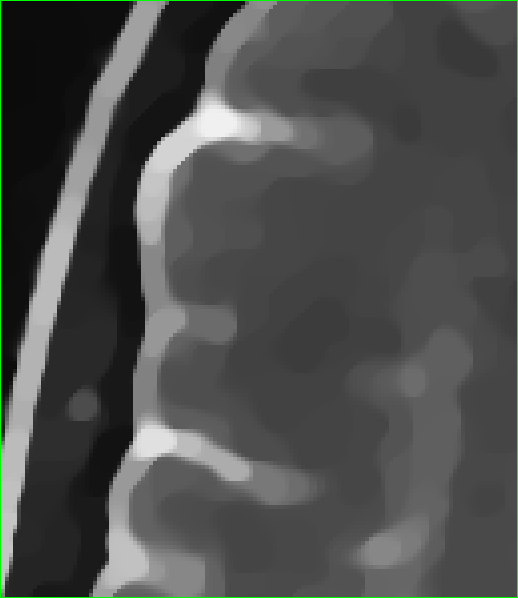} &
		\includegraphics[width=1.1in]{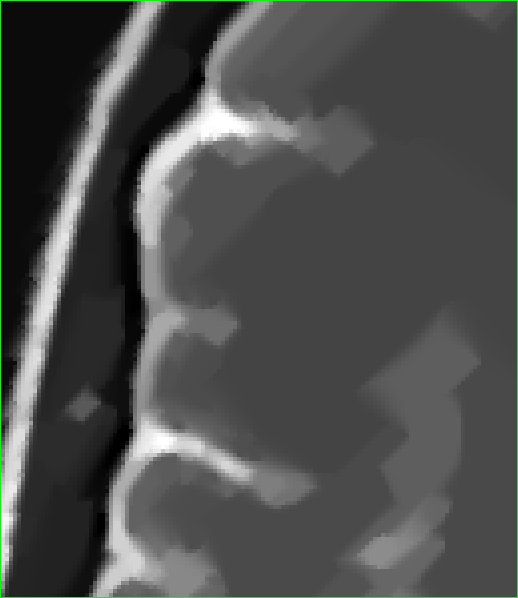}\\
		(e)&(f)&(g)&(h)\\
	\end{tabular}		
	\caption{\emph{First row}. Original test image \texttt{brain} (a), observed image corrupted by Gaussian blur and AWGN with $\sigma=20$ (b), TV reconstruction (c), BLTV reconstruction (d). \emph{Second row}. Close-up(s) of the first row.}
	\label{fig:data}
\end{figure}

\begin{figure}[tbh]
	\center
	\begin{tabular}{cccc}
		\includegraphics[width=1.1in]{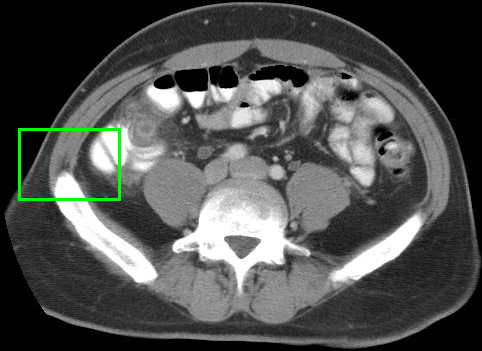} &
		\includegraphics[width=1.1in]{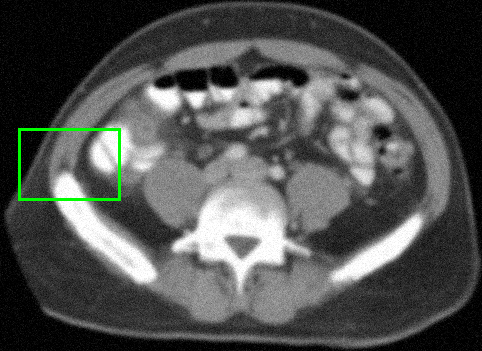} &
		\includegraphics[width=1.1in]{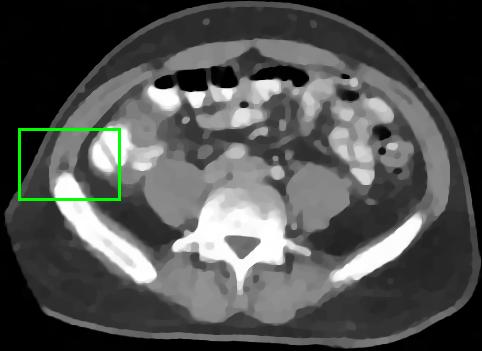} &
		\includegraphics[width=1.1in]{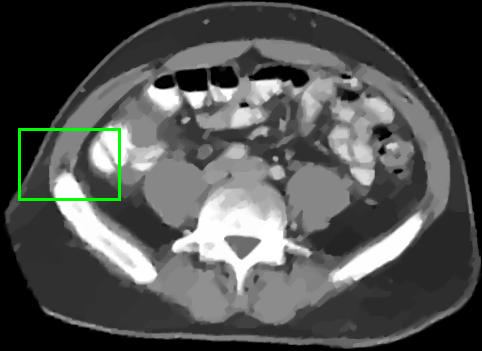}\\
		(a)&(b)&(c)&(d)\\
		\includegraphics[width=1.1in]{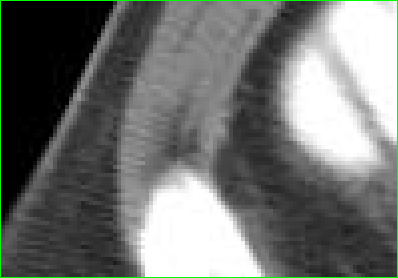} &
		\includegraphics[width=1.1in]{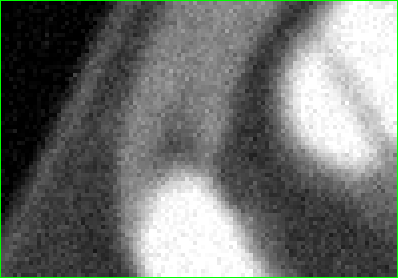} &
		\includegraphics[width=1.1in]{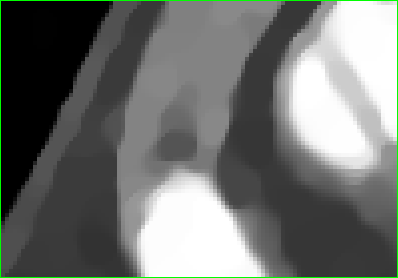} &
		\includegraphics[width=1.1in]{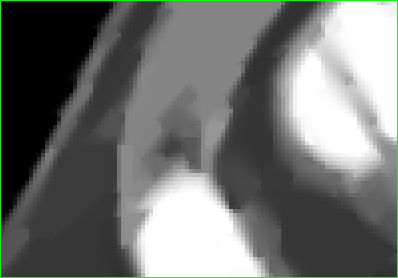}\\
		(e)&(f)&(g)&(h)\\
	\end{tabular}		
	\caption{\emph{First row}. Original test image \texttt{abdomen} (a), observed image corrupted by Gaussian blur and AWGN with $\sigma=10$ (b), TV reconstruction (c), BLTV reconstruction (d). \emph{Second row}. Close-up(s) of the first row.}
	\label{fig:data2}
\end{figure}

We consider the test image \texttt{brain} in Fig.\ref{fig:data} (a) (570 $\times$ 430) and the test image \texttt{abdomen} in Fig.\ref{fig:data2} (a) (350 $\times$ 480) with pixel values between 0 and 255, synthetically corrupted by space-invariant  blur with Gaussian kernel of parameters \verb|band| = 9, \verb|sigma| = 2, and by AWGN of different levels $\sigma \in \{10,20\}$ - see, e.g., Fig.\ref{fig:data} (b) and Fig.\ref{fig:data2} (b). The parameter maps are computed at the beginning starting from the observed image $g$ and then updated every 300 iterations based on the current iterate. The radius of the neighborhoods used for the local parameter estimation has been set equal to 8 and 5 for the \texttt{brain} and \texttt{abdomen} test images, respectively. The reconstructions of \texttt{brain} for $\sigma=20$ and of \texttt{abdomen} for $\sigma=10$ via TV-L$_2$ and BLTV-L$_2$ models are shown in Figs.\ref{fig:data}(c)-(d) and Figs.\ref{fig:data2}(c)-(d), respectively.

\begin{figure}[tbh]
	\center
	\begin{tabular}{ccc}
		\includegraphics[width=1.45in]{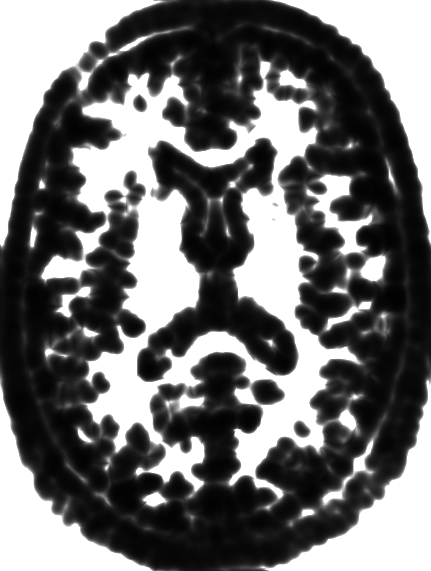} &
		\includegraphics[width=1.45in]{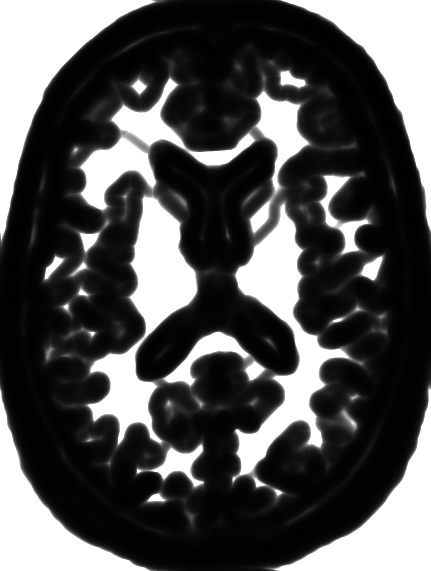} &
		\includegraphics[width=1.45in]{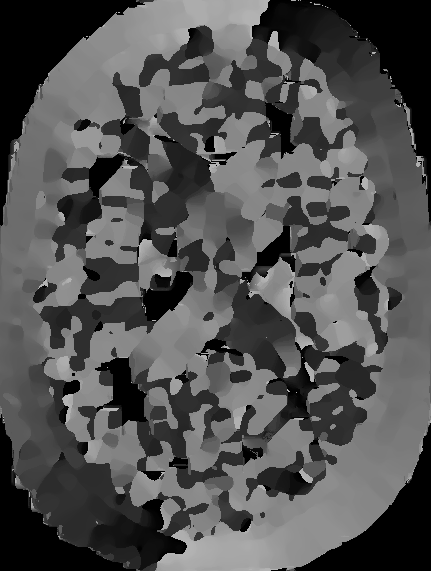}\\
		$\lambda^{(1)}$&$\lambda^{(2)}$&$\theta$\\
		\end{tabular}		
	\caption{Final parameter maps for \texttt{brain} test image.}
	\label{fig:maps}
\end{figure}

\begin{figure}[tbh]
	\center
	\begin{tabular}{ccc}
		\includegraphics[width=1.45in]{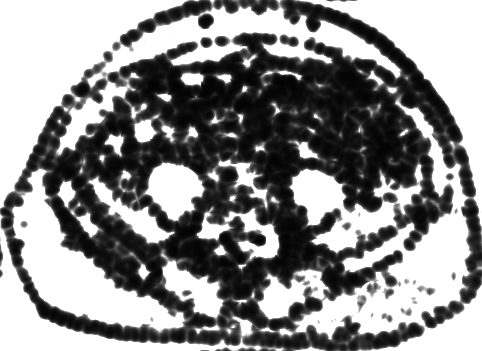} &
		\includegraphics[width=1.45in]{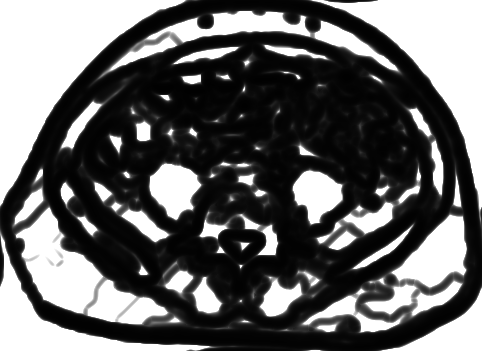} &
		\includegraphics[width=1.45in]{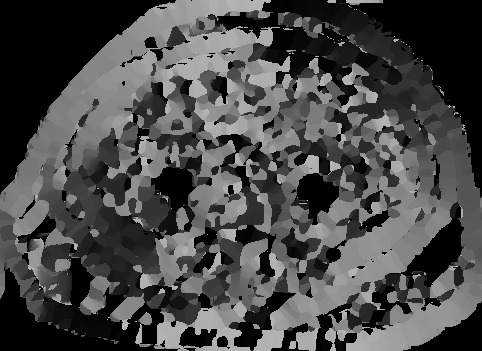}\\
		$\lambda^{(1)}$&$\lambda^{(2)}$&$\theta$\\
	\end{tabular}		
	\caption{Final parameter maps for \texttt{abdomen} test image.}
	\label{fig:maps2}
\end{figure}

From a visual inspection, the restoration via BLTV-L$_2$ seems to be more neat and less cartooned than the TV reconstructions. As reported in Tab.\ref{tab:1}, the ISNR and SSIM values for the two test images and for different noise levels obtained by the BLTV-L$_2$ model outperform the ones reached by the TV-L$_2$ model.\\
The final parameter maps computed by BLTV-L$_2$ are shown in Figs. \ref{fig:maps}-\ref{fig:maps2}. 

	
%
\begin{table}[!t]
	\vspace{0.3cm}
	\centering
	\begin{tabular}{l|cc|cc|cc|cc}
	\hline\hline
&\multicolumn{4}{c}{\textbf{\texttt{brain}}}&\multicolumn{4}{c}{\textbf{\texttt{abdomen}}}\\
		\hline\hline
		&\multicolumn{2}{c}{$\sigma=10$}&\multicolumn{2}{c}{$\sigma=20$}&\multicolumn{2}{c}{$\sigma=10$}&\multicolumn{2}{c}{$\sigma=20$}\\
		\hline\noalign{\smallskip}
		&TV-L$_2$&BLTV-L$_2$&TV-L$_2$&BLTV-L$_2$&TV-L$_2$&BLTV-L$_2$&TV-L$_2$&BLTV-L$_2$\\
		\hline
		ISNR&4.27&\textbf{5.52}&5.00&\textbf{6.67}&3.76&\textbf{4.66}&6.10&\textbf{6.77}\\
		SSIM&0.87&\textbf{0.88}&0.83&\textbf{0.85}&0.78&\textbf{0.80}&0.74&\textbf{0.76}\\
		\noalign{\smallskip}\hline\noalign{\smallskip}
	\end{tabular}
	\caption{Maximum ISNR/SSIM values achieved by TV-L$_2$ and BLTV-L$_2$ on \texttt{brain} and \texttt{abdomen} test images corrupted by AWGN of two different levels.}
	\label{tab:1}
\end{table}

\emph{Computational times.} We tested the joint parameter estimation + reconstruction model on a standard laptop with inbuilt MATLAB software, version 2016b.
As far as the ML parameter estimation of the parameter maps procedure is concerned,
 we notice that the update \eqref{eq:updm} is explicit, thus very cheap, whereas the computation of $\theta^*$ in \eqref{pb:min_theta} requires the solution of an optimisation problem. We solve the problem by line-searching upon a suitable discretization of the parameter space. For Fig. \ref{fig:data}, the ML parameter estimation procedure took 8.45 secs.

The ADMM algorithmic sub-steps with automatic parameter update every 300 iterations computes the numerical solution in 381 secs for the high-resolution image in Fig. \ref{fig:data}. A possible way to accelerate the speed of the algorithm would be the computation the parameter maps only in terms of the given image and not along the iterations, although of course that would render a less accurate result.

\color{black}

\vspace{-0.5cm}

\section{Conclusions and future works}

We presented a new space-variant anisotropic regularization term for image restoration based on the \emph{a priori} statistical assumption that the gradients of the unknown target image distribute locally according to space-variant bivariate Laplace distributions. The high flexibility of the proposed regularizer together with the presented ML parameters estimation procedure and ADMM-based minimization algorithm yield a very effective and efficient approach. Preliminary experiments on images corrupted by blur and AWGN strongly indicate that the proposed variational model achieves high-quality restorations and, in particular, outperforms by far the results obtained by classical TV-regularized restoration models. 
Coupling the proposed regularizer with other fidelity terms suitable for dealing with noises other than Gaussian - such as, e.g., Laplace, Poisson and mixed Poisson-Gaussian \cite{MIXED} - is a matter being studied.

\vspace{-0.5cm}

\end{document}